\documentclass[12pt,,letterpaper]{JHEP3}
\usepackage{graphics}
\usepackage{epsfig}
\usepackage{slashed}

\title{ Holographic superconductor with hidden Fermi surfaces}

\author{ZhongYing Fan\\
Department of Physics, Beijing Normal University, 100875 Beijing,
China\\
\email{zhyingfan@gmail.com}
}

\abstract{In this paper, we investigate a holographic model of superconductor with hidden Fermi surfaces, which was defined by the logarithmic violation of area law of entanglement entropy. We works in fully back-reacted background using standard Einstein-Maxwell-Dilaton action with additional complex scalar filed which was charged under the Maxwell field. Particularly, we analyze the behavior of entanglement entropy during the phase transition. At the critical point, the finite part of the entanglement entropy has a discontinuity of slope and tends to a lower value in the superconducting phase all the way down to the zero temperature limit, indicating the reorganization of degrees of freedom of the system across the phase transition.   }

\keywords{AdS/CFT correspondence, hidden Fermi surfaces, holographic superconductor, condensed-matter theory}
\preprint{}
\begin{document}

\section{Introduction}
Ads/CFT correspondence provides a powerful tool to study the low energy physics in condensed matter systems. The strongly correlated theories in the boundary are mapped to the weakly coupled gravity theories in the bulk. The bulk geometry contains the full information of the boundary systems at all scales such that the dual theory can be systematically analyzed, without additional assumptions and ambiguities. Typical examples have been given in recent years, such as holographic model of superconductors, strange metal, entanglement entropy and fluid turbulence\cite{1,2,3,4,5,HL}.

In order to describe the condensed-matter theories which usually lives in very low energy, people have generalized the correspondence to non-relativistic cases\cite{6,7,8}. The bulk geometry is characterized by dynamic exponent $z$ and hyperscaling violation $\theta$ in the deep interior, which reads
\begin{equation} ds^2=R^2[-\frac{dt^2}{r^{2m}}+g_0r^{2n}dr^2+\frac{dx_i^2}{r^2}] \label{1}\ , \end{equation}
where $R$ is AdS radius, $g_0$ is a positive constant, $i=1,2,...,d$ denotes the spatial dimension, the geometry has been parameterized by two constants $m$ and $n$ which are related to $z$ and $\theta$ by
\begin{equation} z =\frac{m+n+1}{n+2}\ ,\quad   \theta =\frac{n+1}{n+2}\cdot d\ .        \label{2}\end{equation}
Note that when $n=-2$, $z\rightarrow \infty,\   \theta\rightarrow -\infty$, corresponding to a class of spacetime conformally related to $AdS_2\times R_d$ if $z/\theta$ is fixed to be a constant\cite{9}. The metric scales as
\begin{equation}
t \rightarrow \lambda^z t,\ x_i\rightarrow \lambda x_i,\ r\rightarrow \lambda^{(d-\theta)/d} r,\ ds\rightarrow \lambda ^{\theta/d} ds\ .\label{3}\end{equation}  Clearly, the scale invariance is broken for non-zero hyperscaling violation which appears in general below some non-trivial dimensional scale in the dual field theory. The elegant observation given by N.Ogawa et al\cite{10} is that for $\theta=d-1$, the entanglement entropy using holographic prescription\cite{4,12} depends logarithmically on the length of the subsystem by taking the IR limit $\ell\gg r_F$, where $r_F$ is a typical scale in the bulk which was explained to be the inverse of Fermi momentum. In condensed-matter theory, it has been argued that the logarithmic behavior of entanglement entropy shows the existence of Fermi surfaces in above limit\cite{13,14}. To be specific, for a strip subsystem defined by
\begin{equation} A=\{(x_1,x_2,...,x_d)|-\frac{\ell}{2}\leq x_1 \leq \frac{\ell}{2},0\leq x_2,x_3,...,x_d \leq L\},\label{strip} \end{equation}
the entanglement entropy $S_A$ will be substantially modified when $\ell$ is sufficiently large. It behaves like
\begin{equation} S_A=\gamma \frac{L^{d-1}}{\epsilon^{d-1}}+\eta L^{d-1} k_F^{d-1} log(\ell k_F)+O(\ell^0). \label{hee}\end{equation}
where $\gamma$ and $\eta$ are positive constants, $\epsilon$ is the UV cut off, $k_F$ is the Fermi momentum or the average of Fermi momentums when many Fermi surfaces exist.

Thus, the Fermi surfaces can be consistently defined by the logarithmic violation of entanglement entropy in a holographic version for theories with gravity duals. The non-relativistic holographic geometries with hyperscaling violation $\theta=d-1$ characterize the existence of Fermi surfaces of the dual theories.
The Fermi surfaces in this approach are called hidden since there are no explicit fermions in the bulk. From holographic dictionaries, the bulk fermions corresponds to the gauge invariant fermion operators of the boundary theories while the elementary fermion operators with no gauge invariance lacks a holographic description. 

Using the peculiar behavior of entanglement entropy, it becomes possible to characterize the physical properties of these non-gauge invariant operators in holography. Related works appear in \cite{9,15,16}. In this paper, we will study a holographic model of s-wave superconductors for the systems with hidden Fermi surfaces. The order parameter is introduced by a complex scalar field which is charged under the Maxwell field in the full Einstein-Maxwell-Dilaton theory. We in particular elucidate the physical behavior of the entanglement entropy and observe that there is a discontinuous change of the slope of the entanglement entropy at the transition critical point, showing a reorganization of the degrees of freedom of the system. Crossing the critical point, the entanglement entropy always has a lower value in the superconducting phase compared to the normal phase. Intriguingly, the logarithmic behavior is preserved with high precision in the full phase structure of the system.

The remainder of this paper is organized as follows: in section 2, we briefly review the standard Einstein-Maxwell-Dilaton theory which completely realizes the bulk geometries (\ref{1}); in section 3, we construct a holographic model of superconductor, derive equation of motions and impose proper boundary conditions; in section 4, we present the superconducting phase transition by numerically solving the equation of motions;  in section 5, we analyze the behavior of entanglement entropy during the phase transition; in section 6, a short conclusion is given.

\section{Preliminary}
As mentioned above, the general scaling geometries with hyperscaling violation (\ref{1}) appear as the solutions of Einstein-Maxwell-Dilaton (EMD) action. The full EMD theory has been discussed in many literatures such as\cite{7,8,16,17,18}. Here we simply list the model and the solutions. The action reads
\begin{equation} S_{EMD}=\frac{1}{2\kappa^2}\int \mathrm{d}^{d+2}x \sqrt{-g}\ [\mathcal{R}-2(\partial{\phi})^2-V(\phi)-\frac{\kappa^2}{2 e^2}Z(\phi)F^2]. \label{emd}\end{equation}
To obtain eq.(\ref{1}) as the IR solutions of bulk geometry, the equation of motions give
\begin{equation} F^{rt}=F_0r^{(m-n+d)}Z^{-1}(\phi)\ ,\quad  \phi=k_0\log{r}, \quad k_0=\sqrt{\frac d2 (m-n-2)}\ ,\label{fh}\end{equation}

\begin{equation} V(\phi)=-\frac{V_0^2}{R^2} e^{-\beta \phi}\ ,\quad V_0^2=\frac{\delta(m+d-1)}{g_0}\ ,\quad  \beta=\frac{2(n+1)}{k_0}\ , \label{v}\end{equation}

\begin{equation} Z(\phi)=Z_0 e^{\alpha \phi}\ ,\quad \alpha=\frac{2(n+d+1)}{k_0}\ ,\quad Z_0=\frac{\kappa^2F_0^2}{\delta(m-1)}\label{z}.\end{equation}
where $\delta=m+n+d+1$, $F_0$ is a physical parameter which is proportional to the charge density carried by the black-brane. Recall that these are IR solutions which are valid for $1\ll r< r_h$ in a proper coordinate system, where $r_h$ is the location of the horizon. When including the thermal factor in the rr and tt components of the metric (\ref{1}) as $r^{-2m}\rightarrow r^{-2m}h(r), r^{2n}\rightarrow r^{2n} h^{-1}(r)$ with the thermal factor $h(r)=1-(r/r_h)^\delta$, above solutions are naturally extended to the near horizon region without modifications. The stability of the system requires the reality of the dilaton field, implying that $m\geq n+2$ or equivalently $\theta\leq d-1,\ z\geq 1+\theta/d$, compatible with the constraint of null energy conditions\cite{10,16}.

The exponents $\alpha$ and $\beta$ can be further expressed in terms of dynamic exponent $z$ and hyperscaling violation $\theta$. For our interests, we will focus on discussing $d=2$ case in the remainder of this paper. We obtain
\begin{equation} \beta=\frac{\theta}{4-\theta}\alpha,\quad \alpha=\pm\frac{2(4-\theta)}{\sqrt{(2-\theta)(2z-2-\theta)}}. \label{ab} \end{equation}
The IR metric (\ref{1}) should be embedded into a full bulk geometry with asymptotical AdS boundary. This is dual to choosing a complete form of the dilaton potential $V_c(\phi)$ which satisfies eq.(\ref{v}) in the deep IR and $V_c(\phi)=-6/R^2+2M_{\phi}^2 \phi^2$ in the UV limit $r\rightarrow 0$. Here $M_{\phi}$ denotes the dilaton mass. For convenience, we set
\begin{equation} V_c(\phi)=-2\frac{V_0^2}{R^2}\cosh{\beta \phi}.\label{vc} \end{equation}
The UV limit behavior leads to $V_0^2=3$, $M_\phi^2=-\beta^2 V_0^2/2$. The Breitenlohner-Freedman bound requires that $M_\phi^2\geq -9/4$, leading to $\beta^2\leq 3/2$. It should be emphasized that there are many other choices of the full dilaton potential which are different from eq.(\ref{vc}) but share the identical UV and IR behaviors. In general, this will probably give branches of EMD theories which behave much different in the intermediate regions. However, the system we study in this paper encode information of Fermi surfaces in the absence of bulk fermions. The main physic quantity we have interests in is holographic entanglement entropy of the minimal area surface which is dominantly contributed in the deep IR. The contributions from other pieces have been analytically estimated as $O(\Lambda^0)$, $\Lambda=\ell/r_F\gg 1$, compared to the IR contribution\cite{10}. Thus, the behavior of the total entanglement entropy strongly depends on the IR solutions (\ref{v}), indicating that the explicit form of the full potential has little influence on the entanglement entropy. In this sense, we expect that our choice eq.(\ref{vc}) is not a special one. Furthermore, the dialton gauge coupling $Z(\phi)$ can also have a full formula which approaches to the IR solution eq.(\ref{z}) in the bulk interior but we will not consider this complication, simply using eq.(\ref{z}) and setting $Z_0=1$ throughout this paper.

\section{Equation of motions and boundary conditions}
Once dynamic exponent and hypersclaing violation are fixed, every detail of EMD theory will be determined from eq.(\ref{emd}-\ref{vc}). For systems with hidden Fermi surfaces, $\theta=1$, leading to
\begin{equation}\beta=\frac \alpha3,\quad \alpha=\pm\frac{6}{\sqrt{2z-3}}.\label{ab2}  \end{equation}
We will only choose positive values of $\alpha$ and $\beta$ in the following. We find that it is convenient to work by setting $M_\phi^2=-2$ which implies $\beta=2/\sqrt{3}$, $\alpha=6/\sqrt{3}$ and $z=3$. In order to investigate the holographic superconductor with hidden Fermi surfaces, we further introduce a complex scalar field which is charged under the Maxwell field. The total action reads
\begin{equation} S_{tot}=\frac{1}{2\kappa^2}\int \mathrm{d}^{4}x \sqrt{-g}\ [\mathcal{R}-2(\partial{\phi})^2-V_c(\phi)-\frac{\kappa^2}{2 e^2}Z(\phi)F^2-2(|D\psi|^2+M_\psi^2|\psi|^2)]. \label{action} \end{equation}
where $D_\mu=\partial_\mu-i A_\mu$, $A_\mu$ is the gauge potential. $M_\psi^2$ is the mass square of the charged scalar field which will also be set to $M_\psi^2=-2$. Variation of the action with respect to $g^{\mu\nu},\ A_\mu,\ \phi\ \mathrm{and}\ \psi$, we obtain
\begin{eqnarray}
R_{\mu\nu}-\frac 12 Rg_{\mu\nu} &=& [2\partial_\mu \phi \partial_\nu \phi-g_{\mu\nu}((\partial{\phi})^2+\frac{V_c(\phi)}{2})]+[2D_\mu \psi D_\nu^{*}\psi^*-g_{\mu\nu}(|D\psi|^2+M_\psi^2|\psi|^2)]\nonumber\\
                                 &&+\frac{\kappa^2}{e^2}Z(\phi)(g^{\lambda\rho}F_{\mu\lambda}F_{\nu\rho}-\frac 14 g_{\mu\nu}F^2), \label{eom1}
\end{eqnarray}
\begin{equation}\triangledown^{\mu}F_{\mu\nu}=ie^2[\psi^*D_{\nu}\psi-\psi D_{\nu} \psi^*],\label{eom2}  \end{equation}
\begin{equation}\triangledown_\mu \triangledown^\mu \phi-\frac 14 \frac{\partial V_c(\phi)}{\partial \phi}-\frac{\kappa^2}{8e^2}\frac{\partial Z}{\partial \phi}F^2=0,\label{eom3}   \end{equation}
\begin{equation}(\triangledown_\mu-iA_\mu)(\triangledown^\mu-iA^\mu)\psi-M_\psi^2 \psi=0.\label{eom4}  \end{equation}
To proceed, the metric ansatz is taken as
\begin{equation}ds^2=R^2[-f(r)e^{-\chi(r)}dt^2+\frac{dr^2}{r^4 f(r)}+\frac{dx^2+dy^2}{r^2}],\label{metric} \end{equation}
together with
\begin{equation}A=\varphi(r)dt,\quad \phi=\phi(r),\quad \psi=\psi(r).\label{matter}  \end{equation}
From the r component of Maxwell equations, the phase of $\psi$ must be constant. Therefore, without loss of generality, we can take $\psi$ to be real. It is now straightforward to write down a more explicit form of equations of motions as
\begin{equation}\frac{\chi'(r)}{r}=2[\phi'(r)^2+\psi'(r)^2+\frac{e^{\chi(r)}}{r^4f(r)^2}\varphi(r)^2\psi(r)^2]  \label{eoms1}\end{equation}
\begin{eqnarray}\frac{f'(r)}{rf(r)}-\frac{1}{r^2}&=&\phi'(r)^2+\psi'(r)^2+\frac{1}{r^4f(r)}[\frac 12 V_c(\phi)+M_\psi^2\psi(r)^2]\nonumber\\
                                                &&+\frac{e^{\chi(r)}}{2f(r)}Z(\phi)\varphi'(r)^2+\frac{e^{\chi(r)}}{r^4f(r)^2}\varphi(r)^2\psi(r)^2,  \label{eoms2}\end{eqnarray}
\begin{equation}\phi''(r)+[\frac{f'(r)}{f(r)}-\frac{\chi'(r)}{2}]\phi'(r)-\frac 14 [\frac{1}{r^4f(r)}\frac{\partial V_c(\phi)}{\partial \phi}-\frac{e^{\chi(r)}}{f(r)}\frac{\partial Z(\phi)}{\partial \phi}\varphi'(r)^2]=0,  \label{eoms3}\end{equation}
\begin{equation}\varphi''(r)+[\alpha \phi'(r)+\frac{\chi'(r)}{2}]\varphi'(r)-\frac{2Z^{-1}(\phi)}{r^4f(r)}\psi(r)^2\varphi(r)=0,  \label{eoms4}\end{equation}
\begin{equation}\psi''(r)+[\frac{f'(r)}{f(r)}-\frac{\chi'(r)}{2}]\psi'(r)+[\frac{e^{\chi(r)}}{r^4f(r)^2}\varphi(r)^2-\frac{M_\psi^2}{r^4f(r)^2}]\psi(r)=0.  \label{eoms5}\end{equation}
where the prime denotes the derivative with respect to $r$. Note that we have set $R=Z_0=\kappa^2=e^2=1$. There are two scaling symmetries of these equations
\begin{equation}r\rightarrow r/a,\quad (t,x,y)\rightarrow (t/a,x/a,y/a),\quad f(r)\rightarrow a^2 f(r),\quad \varphi(r)\rightarrow a\varphi(r), \label{sc1}\end{equation}
\begin{equation}t\rightarrow t/a,\quad \varphi(r)\rightarrow a\varphi(r),\quad \chi(r)\rightarrow \chi(r)-2\ln{a}. \label{sc2}\end{equation}
The first can be used to set the horizon $r_h=1$ whereas the second can scale the value of $\chi(r)$ to zero in the UV limit $r\rightarrow 0$. To solve the  equations of motions eq.(\ref{eoms1}-\ref{eoms5}), one needs to impose proper boundary conditions at the horizon
\begin{equation} \chi(r)=\chi^{(0)}+\chi^{(1)}(r-1)+\chi^{(2)}(r-1)^2+...,\label{bdy1}\end{equation}
\begin{equation} f(r)=f^{(1)}(r-1)+f^{(2)}(r-1)^2+...,\label{bdy2}\end{equation}
\begin{equation} \phi(r)=\phi^{(0)}+\phi^{(1)}(r-1)+\phi^{(2)}(r-1)^2+...,\label{bdy3}\end{equation}
\begin{equation} \varphi(r)=\varphi^{(1)}(r-1)+\varphi^{(2)}(r-1)^2+...,\label{bdy4}\end{equation}
\begin{equation} \psi(r)=\psi^{(0)}+\psi^{(1)}(r-1)+\psi^{(2)}(r-1)^2+...\label{bdy5}\end{equation}
where the dots denotes higher order terms. Recall that the horizon is defined by $f(r_h)=0$ and the finite norm of the gauge potential requires $\varphi(r_h)=0$. In general, $\chi^{(0)}$ and $\phi^{(0)}$ do not vanish at the horizon but we will simply set them to zero to simplify our numerical code. The complex scalar field also vanishes in the normal phase but has a nonzero value when the system crosses the critical point to the superconducting phase. At the horizon, there are two independent parameters $(\psi^{(0)},\varphi^{(1)})$. Integrating out to the boundary, they are mapped to the physical quantities in the dual theory which will be extracted from the expansion coefficients of the fields at the boundary $r\rightarrow 0$
\begin{equation}\chi(r)=\chi_2 r^2+...,  \label{ep1}\end{equation}
\begin{equation}f(r)=\frac{1}{r^2}+f_1+f_2 r^2+...,  \label{ep2}\end{equation}
\begin{equation}\phi(r)=\phi_1 r+\phi_2 r^2+...,  \label{ep3}\end{equation}
\begin{equation}\varphi(r)=\mu-\rho r+...,  \label{ep4}\end{equation}
\begin{equation}\psi(r)=\psi_1 r+\psi_2 r^2+...  \label{ep5}\end{equation}
where $\mu,\ \rho$ are chemical potential and charge density respectively. To obtain a stable superconducting solution, we need to impose source free condition for the $\psi$ field: $\psi_1=0$ while the other one is treated as the order parameter with dimension 2 and vice versa. That is
\begin{equation} \psi_1=0,\quad \langle O_2\rangle=\psi_2,\label{o1} \end{equation}
\begin{equation} \psi_2=0,\quad \langle O_1\rangle=\psi_1.\label{o2} \end{equation}

\section{Superconducting solution}

\subsection{Phase transition}
Since the equations of motions eq.(\ref{eoms1}-\ref{eoms5}) are highly nonlinear coupled, we will solve them numerically in Mathematica. In fig.\ref{f1}, we present a typical solution in the superconducting phase with the chemical potential $\mu=6.0469$. It is immediately seen that every function plotted in this figure behaves monotonous: start at zero and never have a maximum or minimum value in the intermediate region. This is consistent with the expectation from the equations of motions. Notice that the function $\chi(r)$ shown in this figure has a non-zero value at the boundary $r\rightarrow 0$. To obtain the solution satisfying the asymptotical behavior of eq.(\ref{ep1}), one need to shift the value of $\chi(r)$ to zero using the scaling symmetry eq.(\ref{sc2}). Correspondingly, the time component of the gauge potential is scaled as the way of eq.(\ref{sc2}) such that the physical charge density as well as the chemical potential will be multiply by a factor of $\exp{(\chi(r)/2)}\mid_{r\rightarrow 0}$.
\FIGURE[ht]{\label{f1}
\includegraphics[width=7cm]{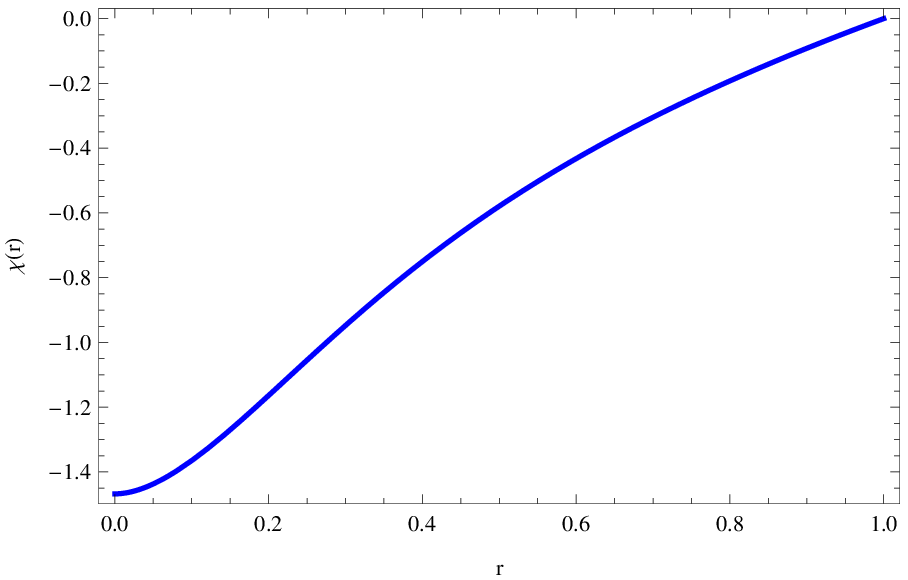}
\includegraphics[width=7cm]{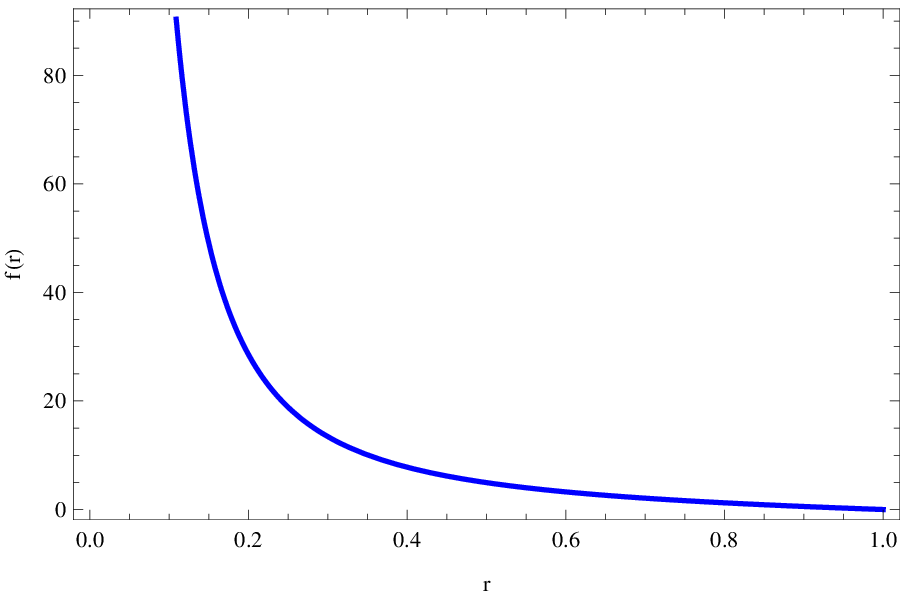}
\includegraphics[width=7cm]{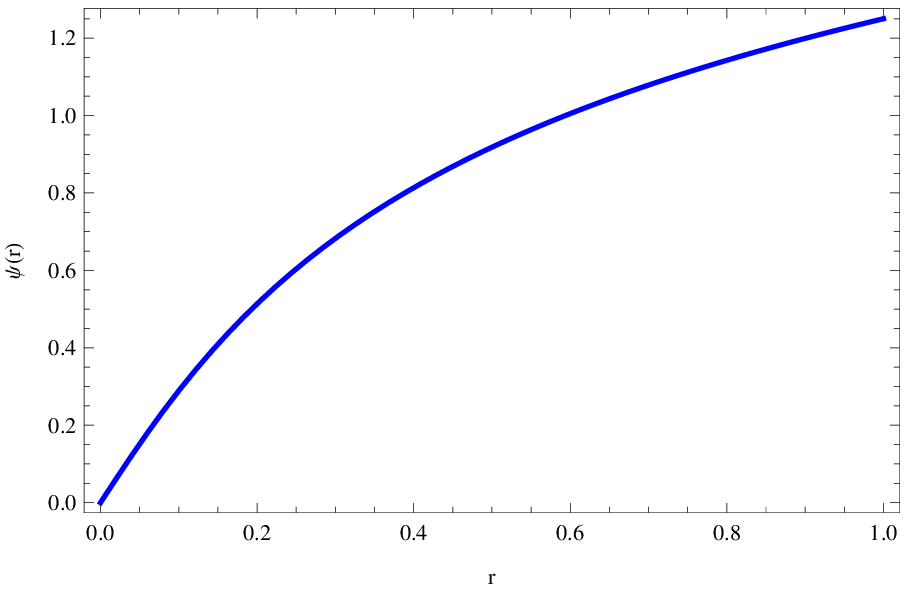}
\includegraphics[width=7cm]{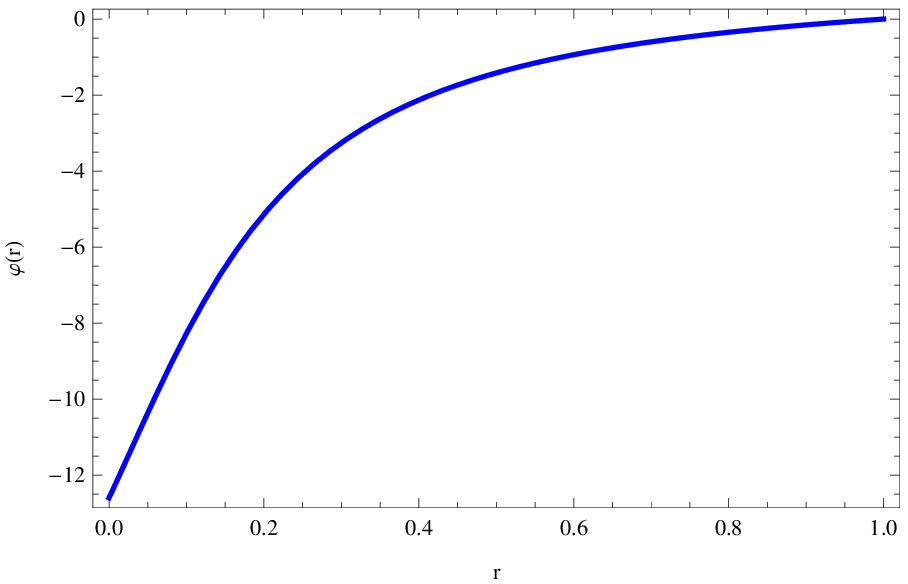}
\caption{A typical solution of the superconductor is given with the chemical potential $\mu=6.0469$. The functions $\chi(r)$ and $\varphi(r)$ have to be furhter scaled as eq.(\ref{sc2}) to obtain a physical solution.}
}
In fig.\ref{f2}, we show the scalar condensates and charge densities as a function of chemical potential. The critical chemical potential is given by $\mu_c=1.1117$ for $\langle O_1 \rangle$ and $\mu_c=12.1756$ for $\langle O_2 \rangle$, respectively.

In both cases, the order parameter is turned on a non-zero value when the chemical potential crosses the transition point and approaches to a constant in the zero temperature limit $T\sim 1/\mu\rightarrow 0$. On the other hand, the charge density increases continuously with the chemical potential, indicating that the phase transition is of second order, which will be further confirmed in the free energy differences as was shown below.

\subsection{Thermodynamics}
Before discussing the entanglement entropy during the phase transition, let's first study the thermodynamical property of the black-brane. The free energy of the system is defined by the on-shell action, up to normalization. The on-shell action reads
\begin{equation}I_{OS}=I_{EH}+I_{GH}+I_{CT},\qquad \mathrm{with}  \label{t1}\end{equation}
\begin{equation}I_{EH}=-S_{tot}=\frac{1}{2\kappa^2}\int_\partial \mathrm{d}^3x\frac{f(r)e^{-\chi(r)/2}}{r}, \label{t2}\end{equation}
\begin{equation}I_{GH}=-\frac{1}{\kappa^2}\int_\partial \mathrm{d}^3x\sqrt{h}(\mathcal{K}-2),  \label{t3}\end{equation}
\begin{equation}I_{CT}=\frac{1}{\kappa^2}\int_\partial \mathrm{d}^3x\sqrt{h}(\phi^2+\psi^2).  \label{t4}\end{equation}
\FIGURE[ht]{\label{f2}
\includegraphics[width=7cm]{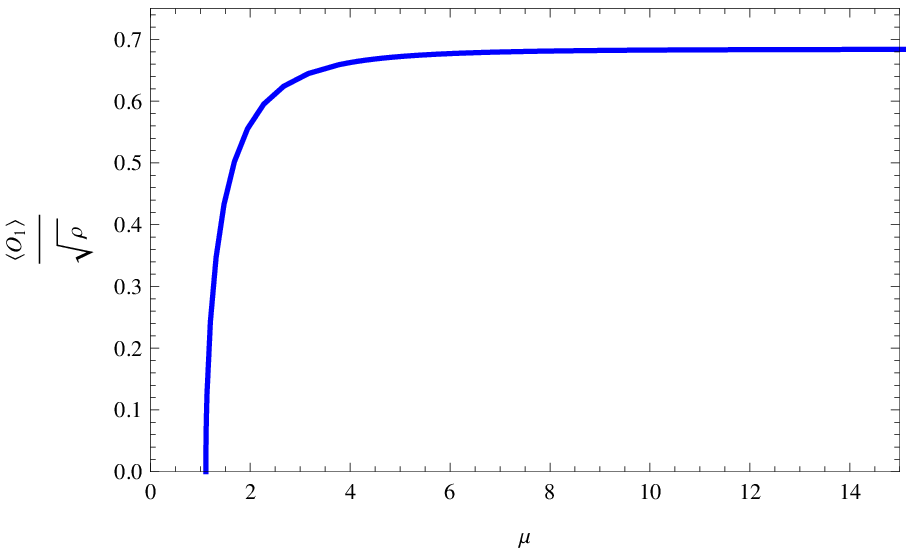}
\includegraphics[width=7cm]{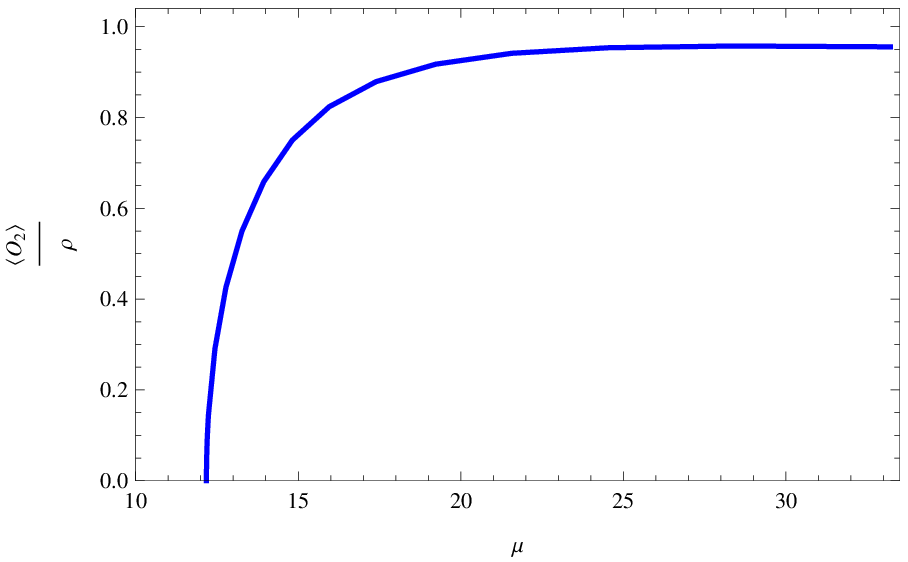}
\includegraphics[width=6.7cm]{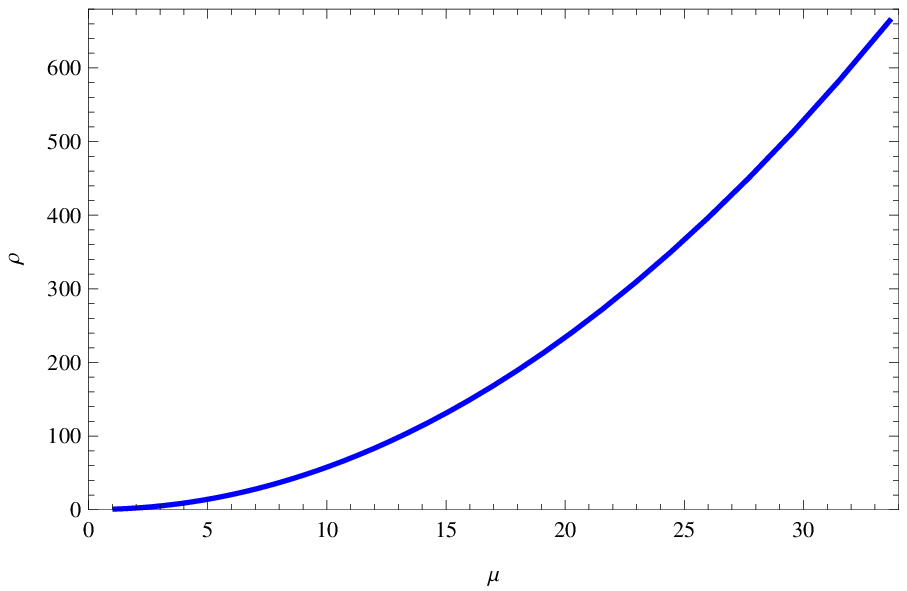}
\includegraphics[width=6.7cm]{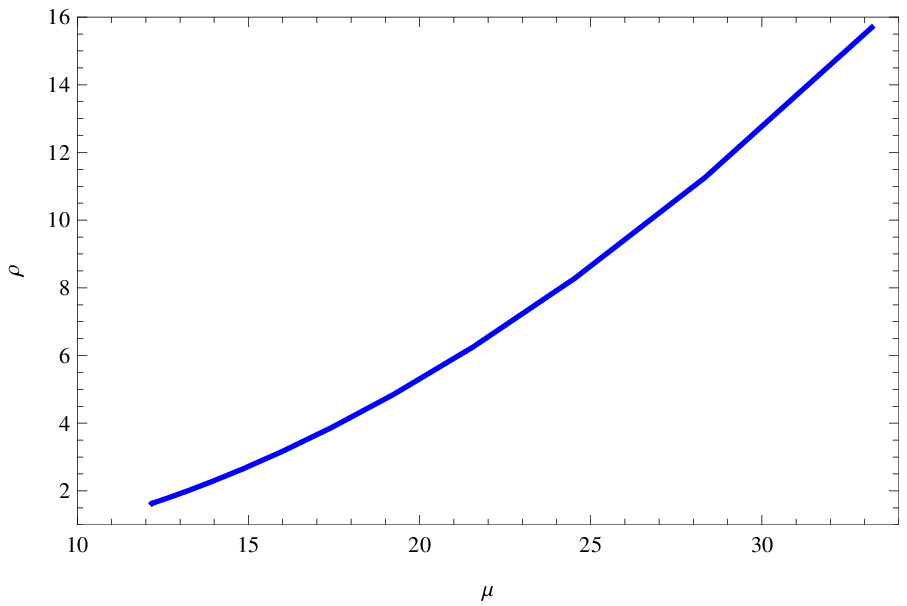}
\caption{The scalar condensate and charge density as a function of chemical potential. The left plots for $\langle O_1\rangle$, the right plots for $\langle O_2\rangle$, respectively.}
}
Here $\partial$ denotes the boundary hypersurface $r=\epsilon\rightarrow 0$, $h$ is the Euclidean version of the determinant of the induced metric and $\mathcal{K}=h^{\mu\nu}\mathcal{K}_{\mu\nu}$, $\mathcal{K}_{\mu\nu}$ is the extrinsic curvature. In eq.(\ref{t2}), the bulk action has been expressed as a total
derivative term by employing Einstein equations of motions. In writing down the counter term action $I_{CT}$ from the matter fields, we have assumed to impose Dirichlet boundary conditions for both $\phi$ and $\psi$ fields at the boundary i.e. fixing $\phi_1$ and $\psi_1$. However, the $\langle O_1 \rangle$ superconductor solution is obtained by fixing $\psi_2$ at the boundary which means that we need to impose Neumann boundary conditions. In this case, the dual counter term of $\psi$ field has a sightly different form
\begin{equation} I_{2}=-\frac{1}{\kappa^2}\int_\partial \mathrm{d}^3x\sqrt{h}(\psi^2+2\psi n^\mu\partial_\mu\psi).  \end{equation}
\FIGURE[ht]{\label{f3}
\includegraphics[width=7cm]{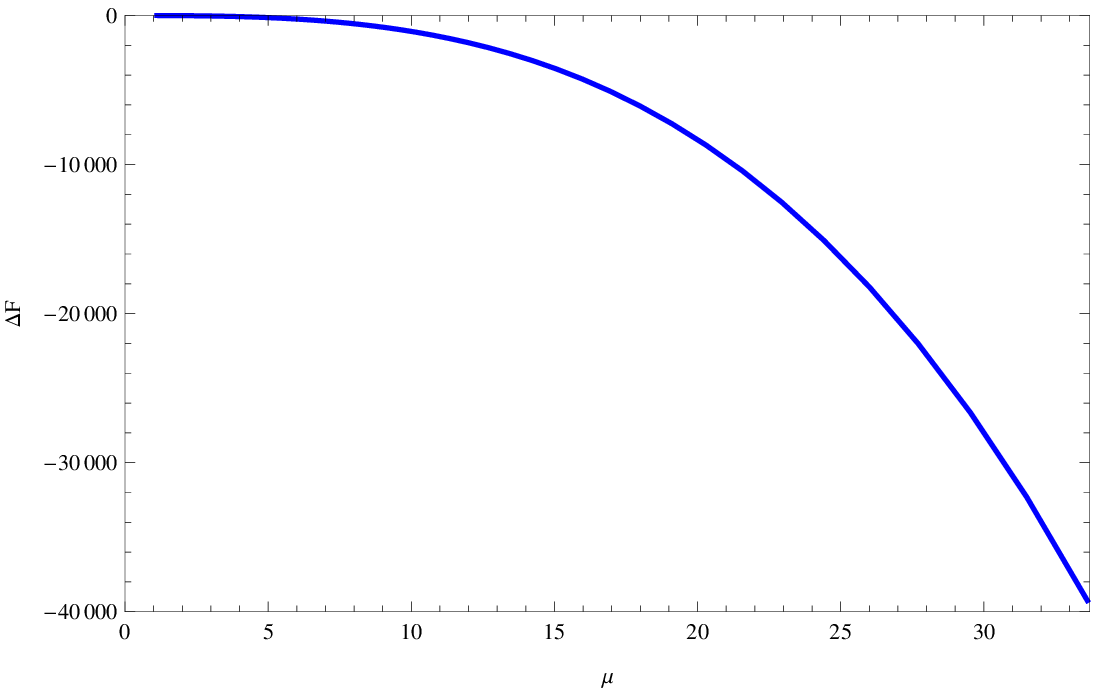}
\includegraphics[width=7cm]{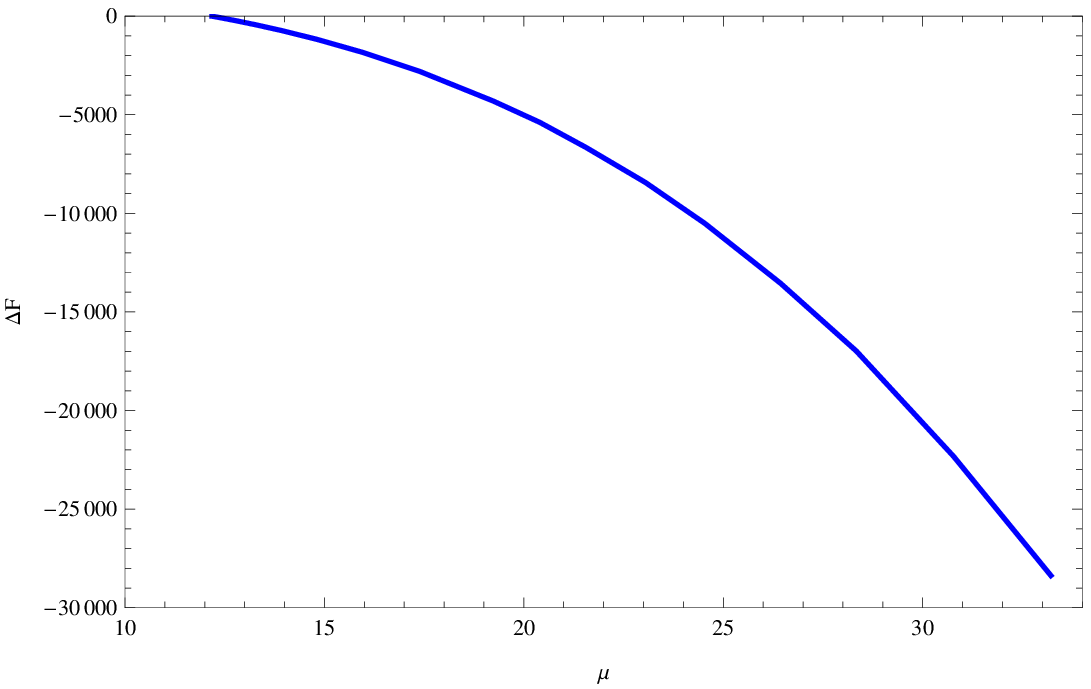}
\caption{The free energy difference as a function of chemical potential. The left plot for $\langle O_1\rangle$, the right plot for $\langle O_2\rangle$, respectively.}
}
Fortunately, due to the source free conditions, the final result of the free energy shares an identical formula for both $\langle O_1 \rangle$ and $\langle O_2 \rangle$ solutions
\begin{equation} F=\frac{2\kappa^2TI_{OS}}{V_2}=f_2+4\phi_1\phi_2.\label{fe} \end{equation}
where $V_2$ denotes the volume factor of the two dimensional space upon which the boundary theory resides. We have defined the free energy as a physical quantity from the on-shell action divided by the volume factor with proper normalisation.

In fig.\ref{f3}, the free energy differences between the superconducting and normal phases are shown as a function of chemical potential. The differences vanish in the normal phase with $\mu<\mu_c$. When the system crosses the critical point to begin superconducting, a non-zero value of the differences is turned on. When $\mu>\mu_c$, the difference monotonously decreases with the increasing of chemical potential, implying that the phase transition is of second order and thermodynamically favored.

\section{Entanglement entropy}
Given the full back-reacted solutions of EMD theory with a charged scalar field, we are ready to study the entanglement entropy during the phase transition. The subsystem $A$ living in the boundary is extended to the minimal area surface $\gamma_A$ in the bulk which satisfies $\partial{\gamma_A}=\partial A$. The entanglement entropy is given by\cite{4,12}
\begin{equation} S_{HEE}=\frac{\kappa^2}{2\pi}S_A. \label{ee1}\end{equation}
where $S_A$ denotes the area of $\gamma_A$. This formula is generally valid for Einstein's gravity with no higher curvature and derivative terms. The shape of the subsystem is free to be chosen. To be specific, we will focus on discussing the strip subsystem defined by eq.(\ref{strip}). The minimal area surface is characterized by $x_1=x_1(r)$ which will be further determined from the extremum condition. It is convenient to first derive the entanglement entropy in a generally static background with rotational and translational invariance. The metric ansatz is taken by
\begin{equation}ds^2=R^2[g_{tt}dt^2+g_{rr}dr^2+\frac{dx_i^2}{r^2}].\label{metric2} \end{equation}
where $g_{tt}, g_{rr}$ are only functions of $r$, $i$ runs over $1, 2, ..., d$. The entanglement entropy of the strip subsystem can be directly deduced as
\begin{equation} S_{HEE}=\frac{\kappa^2}{\pi}R^dL^{d-1}\int_\epsilon^{r_*}\frac{\mathrm{d}r}{r^d}\sqrt{r^2g_{rr}+x_1'(r)^2}.\label{ee2} \end{equation}
Here $\epsilon$ is the UV cut-off, $r_*$ is the turning point of the minimal area surface $\gamma_A$ which will be defined by $x_1'(r_*)\rightarrow \infty$. Variation of the``action" $S_{HEE}$ respect to the ``field" $x_1(r)$, we obtain
\begin{equation}x_1'(r)=\frac{r^d}{r_*^d}\sqrt{\frac{r^2g_{rr}}{1-\frac{r^{2d}}{r_*^{2d}}}}.\label{ee3}  \end{equation}

By simple calculation, we can straightforwardly write down the final results of the entanglement entropy and the width of the subsystem as
\begin{equation} \ell=2\int_0^{r_*}\mathrm{d}r \frac{r^d}{r_*^d}\sqrt{\frac{r^2g_{rr}}{1-\frac{r^{2d}}{r_*^{2d}}}},\label{ee4} \end{equation}
\begin{eqnarray} S_{HEE}&=&\frac{\kappa^2}{\pi}R^dL^{d-1}\int_\epsilon^{r_*}\frac{\mathrm{d}r}{r^d}\sqrt{\frac{r^2g_{rr}}{1-\frac{r^{2d}}{r_*^{2d}}}},\nonumber\\
       &=&\frac{\kappa^2}{\pi}R^dL^{d-1}(\frac{\eta_d}{\epsilon^{d-1}}+s). \label{hee5}\end{eqnarray}
where $\eta_d=2/(d-1)$. The leading divergent term of $S_{HEE}$ agrees with the area law, which is universal for asymptotical AdS background. The physical quantity $s$ which is introduced in the entropy integral eq.(\ref{hee5}) denotes the finite entanglement entropy per unit area,  which is more convenient to be shown in numerics.
\FIGURE[ht]{\label{f4}
\includegraphics[width=7cm]{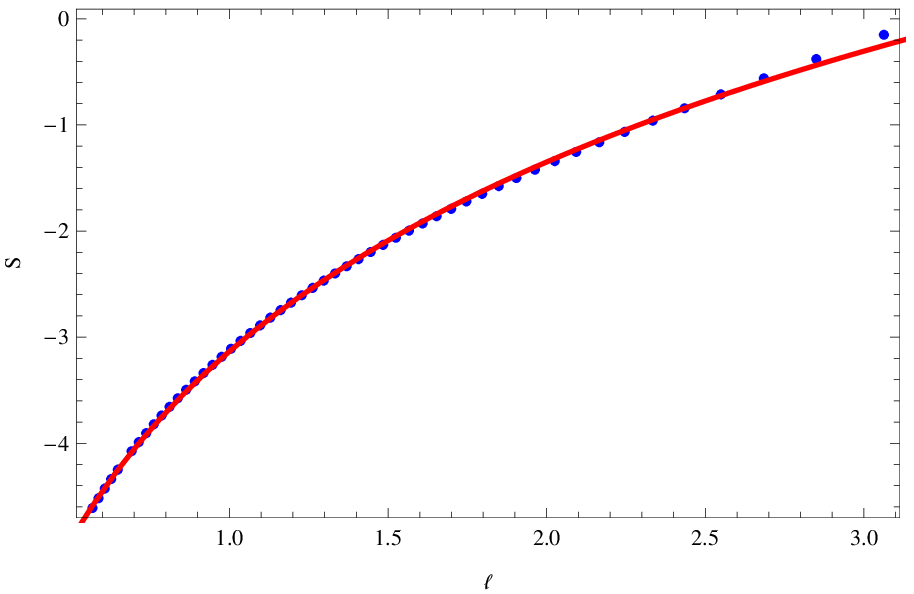}
\includegraphics[width=7cm]{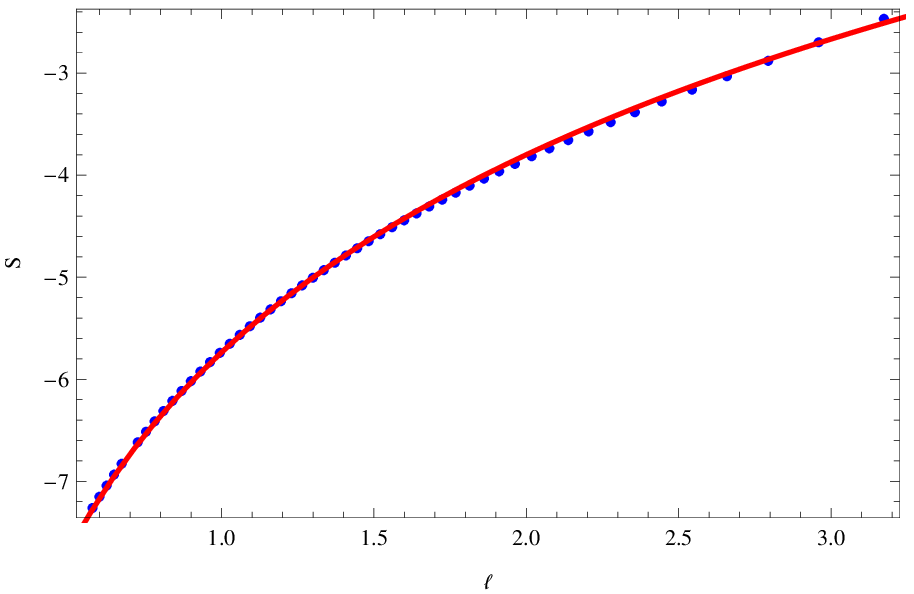}
\includegraphics[width=7cm]{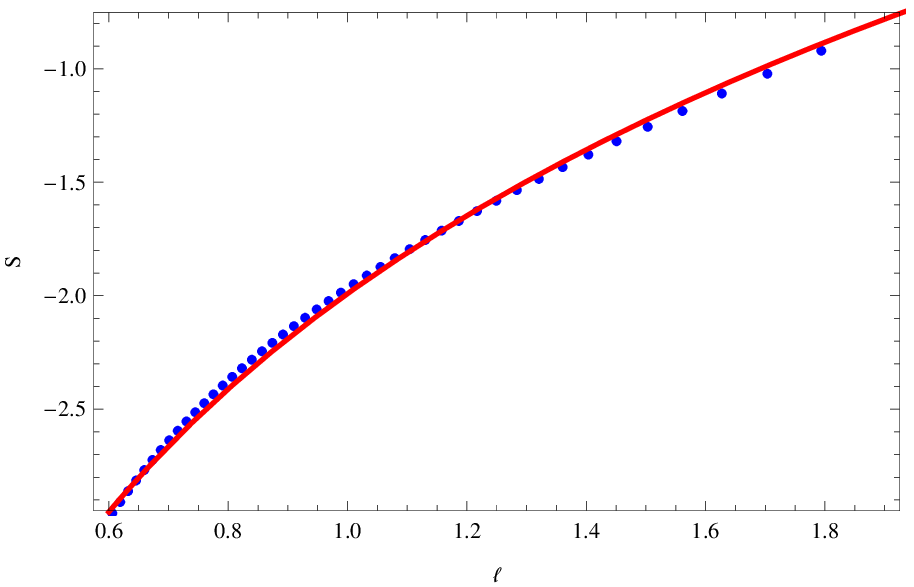}
\includegraphics[width=7cm]{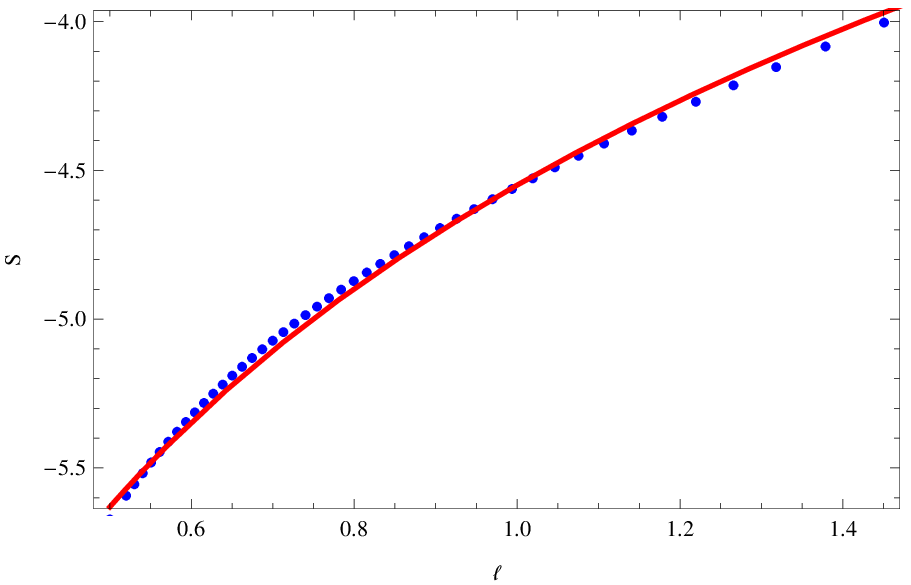}
\caption{The entanglement entropy $s$ as a function of $\ell$ for operator $\langle O_1 \rangle$ with fixed chemical potential. For the normal phase (plots above), $\mu=7.2735\ \mathrm{(left)},\ 12.1756\ (\mathrm{right})$, for the superconduting phase (plots below), $\mu=3.1623\ \mathrm{(left)},\ 8.2083\ (\mathrm{right})$. }
}
Getting back to our superconducting system, we need simply substitute the metric ansatz eq.(\ref{metric}) into eq.(\ref{ee4}-\ref{hee5}), i.e., $d=2$, $g_{rr}=\frac{1}{r^4f(r)}$ . The resulting expressions of $s$ and $\ell$ are integrated in numerical codes. We will focus on studying $\langle O_1 \rangle$ superconductor. The other case,  $\langle O_2 \rangle$ superconductor can be studied in a parallel way. In fig.\ref{f4}, the entanglement entropy $s$ is expressed as a function of the width $\ell$ of the subsystem with fixed chemical potential. The plots above are presented for the normal phase whereas the the plots below are given for the superconducting phase. Here the normal phase means that the charged scalar field $\psi$ is set to zero with no constraints on the chemical potential. For both phases, we take two typical values of the chemical potential to show the relation between $s$ and $\ell$. In every panel of fig.\ref{f4}, we find that the entanglement entropy depends logarithmically on the width of the subsystem. The numerical points can be fitted with high precision as
\begin{equation}s=a\log{\ell}+b.\label{fit1}  \end{equation}
For the normal phase,
\begin{eqnarray} a&=&2.5753,\quad b=-3.1138\quad \mathrm{for}\ \mu=7.2735,\nonumber\\
                 a&=&2.7937,\quad b=-5.7355\quad \mathrm{for}\ \mu=12.1756,\label{fit2}\end{eqnarray}
For the superconductor phase,
\begin{eqnarray} a&=&1.8838,\quad b=-1.9905\quad \mathrm{for}\ \mu=3.1623,\nonumber\\
                 a&=&1.5600,\quad b=-4.5494\quad \mathrm{for}\ \mu=8.2083.\label{fit3}\end{eqnarray}

\FIGURE[ht]{\label{f5}
\includegraphics[width=7cm]{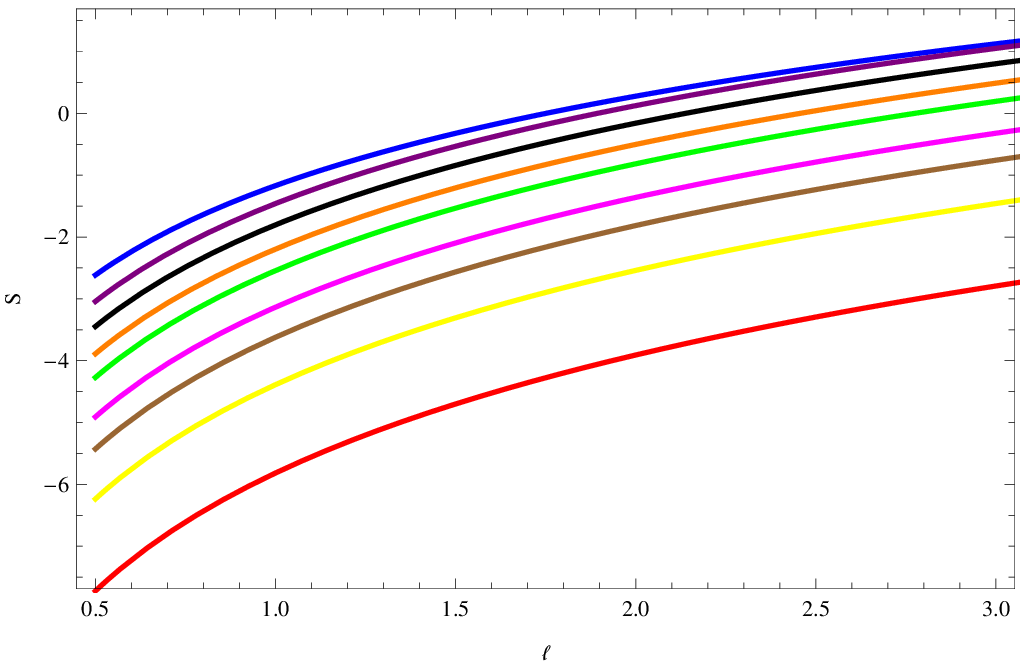}
\includegraphics[width=7cm]{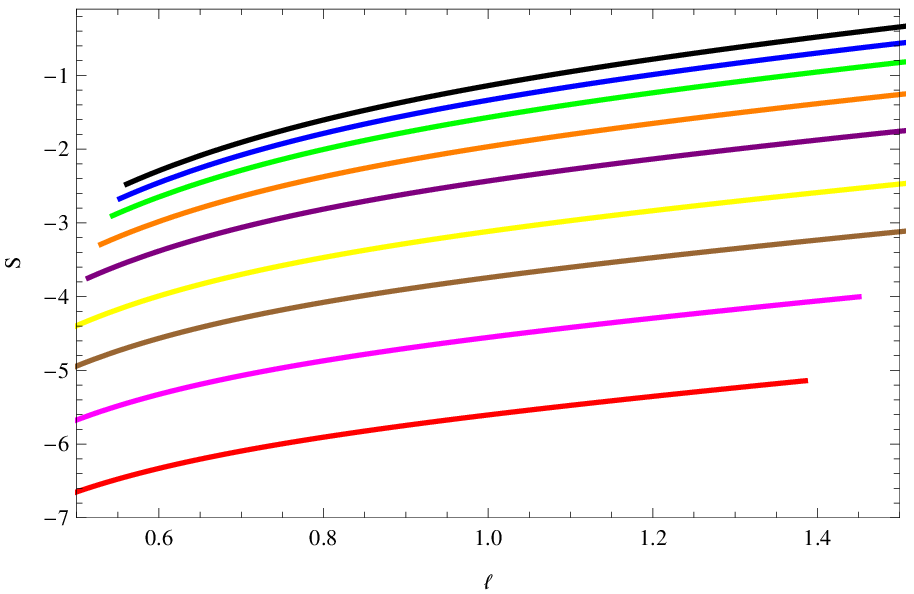}
\caption{The entanglement entropy $s$ as a function of $\ell$ for various chemical potential. In both panels, the chemical potential increases from top to bottom. For the normal phase (left), $\mu=1.0713\ \mathrm{(blue)}$, $\mu=3.1122\ \mathrm{(purple)}$, $\mu=4.1881\ \mathrm{(black)}$, $\mu=5.1988\ \mathrm{(orange)}$, $\mu=6.0068\ \mathrm{(green)}$, $\mu=7.2735\ \mathrm{(magenta)}$, $\mu=8.2553\ \mathrm{(brown)}$, $\mu=9.7285\ \mathrm{(yellow)}$, $\mu=12.3249\ \mathrm{(red)}$. For the superconductor phase (right), $\mu=1.1117\ \mathrm{(black)}$, $\mu=1.6832\ \mathrm{(blue)}$, $\mu=2.2696\ \mathrm{(green)}$, $\mu=3.1623\ \mathrm{(orange)}$, $\mu=4.1307\ \mathrm{(purple)}$, $\mu=5.4822\ \mathrm{(yellow)}$, $\mu=6.6824\ \mathrm{(brown)}$, $\mu=8.2083\ \mathrm{(magenta)}$, $\mu=10.1637\ \mathrm{(red)}$. }
}
The logarithmical behavior of the entanglement entropy presented above is consistent with the analytical results given in \cite{10}. The fact that the entanglement entropy behaves logarithmical in the superconducting phase implies that the solution we obtain is a superconductor with hidden Fermi surfaces.

We first claim that the logarithmical relation between $s$ and $\ell$ is universal for our system with hyperscaling violation $\theta=1$ toward the zero temperature limit $T\sim 1/\mu\rightarrow 0$. In fig.\ref{f5}, we show more lines with various values of chemical potential for both normal and superconductor phases. Every line in the figure runs logarithmically. In both panels, the chemical potential increases from top to bottom. It is immediately seen that the entanglement entropy decreases with chemical potential for fixed $\ell$.

In fig.\ref{f6}, we further show the entanglement entropy as a function of the chemical potential with fixed $\ell=1.2$. The physical curve is always determined by choosing the points with lowest entropy. We observe that there is a discontinuity in the slope of the entanglement entropy at the transition chemical potential (indicated by the red point). The slope is expected to be negative since increasing chemical potential should reduce the number of degrees of freedom as well as the entanglement entropy. The discontinuous change in the slope at the critical point indicates a significant reorganization of the degrees of freedom of the system. Due to the condensate generated in the transition, it is naturally to be expected that there is a large reduction in the number of degrees of freedom as well. From these discussions, we argue that the entanglement entropy provides an independent probe of the phase structure of the superconductor.
\FIGURE[ht]{\label{f6}
\includegraphics[width=9cm]{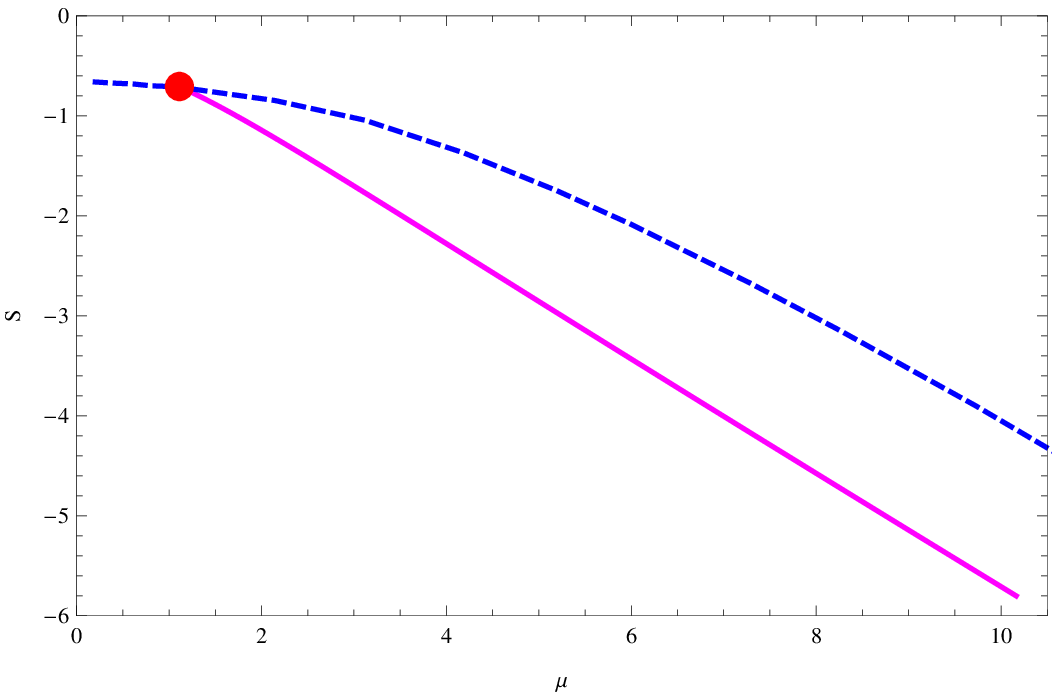}
\caption{The entanglement entropy $s$ as a function of chemical potential $\mu$ for fixed $\ell=1.2$. The dashed blue line denotes the normal phase while the magenta line denotes the superconductor phase. The critical point is denoted by the red point. Trace the physical curve by always choosing the lowest entropy for a given chemical potential.}
}
\section{Conclusions}
In this paper, we investigate a holographic model of s-wave superconductor with hidden Fermi surfaces. We employ the standard Einstein-Maxwell-Dilaton action with additional complex scalar field charged under the gauge field. Varying the chemical potential, we observe a second order phase transition at a critical point.

In particular, the behavior of the entanglement entropy is analyzed with great detail. For systems with hidden Fermi surfaces, the entanglement entropy depends logarithmically on the width of the subsystem. This is a consistent definition for Fermi surfaces in the absence of bulk fermions in the gravity background. By numerically solving the equations of motions, we find it is valid for both normal and superconductor phases. The logarithmical behavior of the entanglement entropy is quite universal without constraints on the chemical potential toward the zero temperature limit. On the other hand, when fixing $\ell$, a discontinuity of the slope of entanglement entropy is observed at the transition point, signifying a reorganization of degrees of freedom of the system during the phase transition. When crossing the critical point, the entanglement entropy always tends to a lower value in the superconducting phase in contrast with the normal phase. These peculiar properties show the utility of the entanglement entropy to independently probe the phase structure of the system.

\section{Acknowledgments}
I would like to thank Professor Sije Gao for his valuable comments on the first version of this manuscript. This work is supported by NSFC Grants NO. 11375026, NO. 11235003 and NCET-12-0054.

\end{document}